\documentclass[aps,superscriptaddress,twocolumn,showpacs,preprintnumbers,amsmath,amssymb]{revtex4}
\usepackage[dvips]{graphicx}
\usepackage{bm}
\usepackage{color}

\newcommand{\comment}[1]{}

\begin{document}
\title{
Remnant Break-up and Muon Production in Cosmic Ray Air Showers
}

\medskip

\author{Hans-Joachim Drescher}
\affiliation{
Frankfurt Institute for Advanced Studies (FIAS),
Johann Wolfgang Goethe-Universit\"at,
Max-von-Laue-Str.~1, 60438  Frankfurt am Main, Germany
}

\begin{abstract}
We discuss the relation between remnant fragmentation in inelastic
high-energy hadronic interactions and muon production in extensive
cosmic ray air showers.  Using a newly developed tool, a simple and
flexible hadronic event generator, we analyze the forward region of
hadronic interactions. We show that measurements of the Feynman-$x$
distribution in the beam fragmentation region at LHCf will be key to
understanding muon production in air showers quantitatively.
\end{abstract}

\maketitle
\section{Introduction}

Muon production in cosmic-ray induced air showers has been of great
interest recently.  The number of muons observed by the Auger collaboration
exceeds the one predicted by popular hadronic interaction models by as
much as 50\% \cite{Schmidt:2007gy}.

A possible mechanism for more muon production in air-shower models has
been proposed by the \textsc{EPOS} model \cite{Pierog:2006qv}, which
describes successfully a large variety of RHIC data. In that model,
the production of most hadron species (in particular, of baryons)
relative to neutral pions is enhanced. Thus, a larger fraction of the
shower energy remains in the hadronic channel which, in turn, leads to
a higher number of muons than in other hadronic air shower models such
as \textsc{QGSjet-II} \cite{qgsjet} and \textsc{Sibyll}\cite{Sibyll}.

In this paper we point to another mechanism that influences muon
production in air showers significantly, namely the treatment of the
remnant of the projectile in a hadronic interaction. The break-up of
the hadronic remnant determines the amount of energy remaining in the
hadronic channel of the air shower (by producing baryons or charged
mesons) relative to that ``lost'' to the electromagnetic part due to
production of neutral pions. Remnant break-up shows up in the very
forward region (large Feynman-$x_F$) of collider experiments and
therefore measurements of the LHCf collaboration \cite{LHCf} will be
important to understand muon production in air showers.

A generic difficulty in studying hadronic interactions for air showers
is the fact that one always compares inherently different Monte-Carlo
event generators. As typical result of detailed comparisons one one
has model A which produces e.g. more muons than model B. One can then
compare generic properties of the underlying hadronic models, like
multiplicity, baryon production, strangeness or charm production,
forward scattering and try to argue for an observed effect. This
method remains always very qualitative, since one cannot disentangle
the importance of the different contributions and one is always left
with the uncertainty that other (possibly neglected) effects 
are actually also important.

In order to relate a specific property of air showers (e.g.\ muon
number) unambiguously to a specific feature of a given hadronic
interaction model (e.g.\ remnant break-up) we developed a new hadronic
interaction model. The goal is to keep the model as simple and
flexible as possible. We then study air shower properties by changing
only one parameter at a time. To investigate the influence of the
remnant break-up on muon production we explore different treatments of
the remnant but do not vary anything else. Likewise, the effects from
baryon production can be studied by exclusively varying the diquark
probability in the string fragmentation.

\section{The tool, the pQCD event generator Picco}

The new tool, \textsc{Picco} (pQCD Interaction Code for COsmics), is a
standard event generator. Details will be published elsewhere, here we
just outline the basic features.  The interaction is modeled by a
superposition of a soft and a semi-hard Pomeron. The Pomeron
parameters are fit to reproduce the total and elastic cross section
for hadron-hadron scattering. Hard scattering is performed by
\textsc{Pythia} \cite{Pythia}, and the partons are mapped onto
strings. Each Pomeron gives exactly 2 strings each of which connects a
quark-antiquark pair scattered out of the interacting particles.  The
longitudinal momentum distribution of the string-ends is the same as
in other models, $dx/\sqrt{x}$ for each (anti)-quark. The remaining
particle, the remnant, has a $x^{1.5}dx$ distribution in the case of
baryons, versus $x^{-0.5}dx$ for mesons \cite{Kaidalov:1983vn}.
Fragmentation of the strings is done via
the Lund fragmentation scheme as implemented in \textsc{Pythia}.

The treatment of the remnant is important for air shower
applications. With a given probability $p_{\rm ex}$, the remnant can
be in an excited state with an invariant mass $M$; the distribution
for $M$ is $dM^2/M^2$. To model the fragmentation of such an excited
remnant state, it could be viewed, for example, as a quark matter
droplet or as a simple string.  In our model we choose the string
approach. If the remnant has a remaining light-cone momentum fraction
$x^+$ (which is the momentum fraction of one string-end), the
corresponding momentum fraction of the other string-end is given by
$x^-=M^2/x^+$.  Remnant strings may have different fragmentation
parameters in order to enhance forward baryon or strangeness
production.

The remnant treatment outlined above is not new, it has been used in
many models: \textsc{QGSjet-98/01}~\cite{Kalmykov:1997te},
\textsc{QGSjet-II}~\cite{qgsjet}, the versions
\textsc{neXus~2}~\cite{Drescher:2000ha} and
\textsc{neXus~3}~\cite{Werner:2003hf} and \textsc{EPOS}
\cite{Pierog:2006qv}. In other models, the remnant is modelled by
attaching one of the strings resulting from Pomeron exchange to a
diquark. While this approach has some drawbacks which will be outlined
below, it is also theoretically less appealing, because one of the
interactions (the one including the diquark) is preferred over the
other ones. By separating the remnant, all interactions are treated on
equal footing.

\section{Soft remnant break-up}

\begin{figure}[tb]
\begin{center}
\includegraphics[width=0.9\columnwidth]{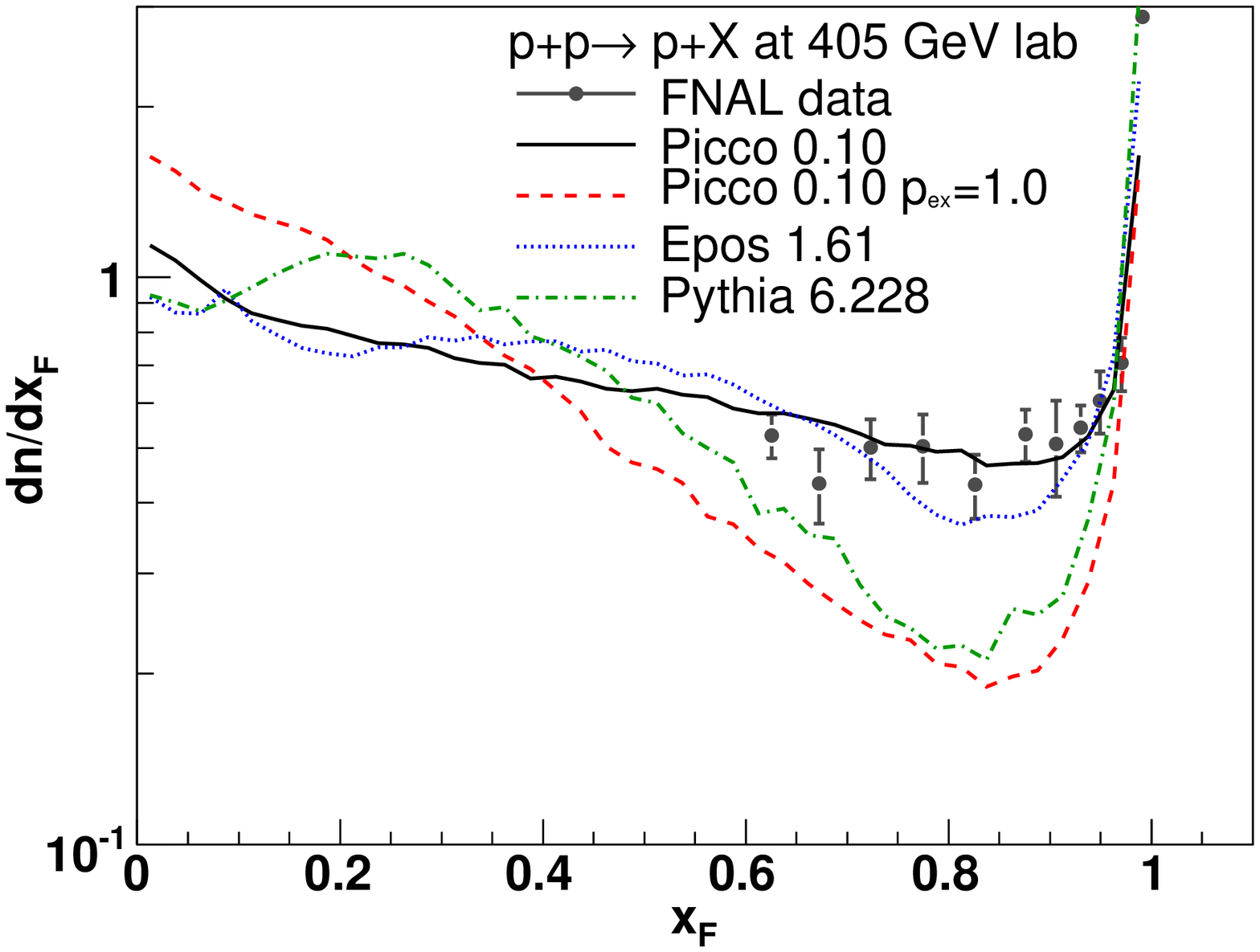}
\includegraphics[width=0.9\columnwidth]{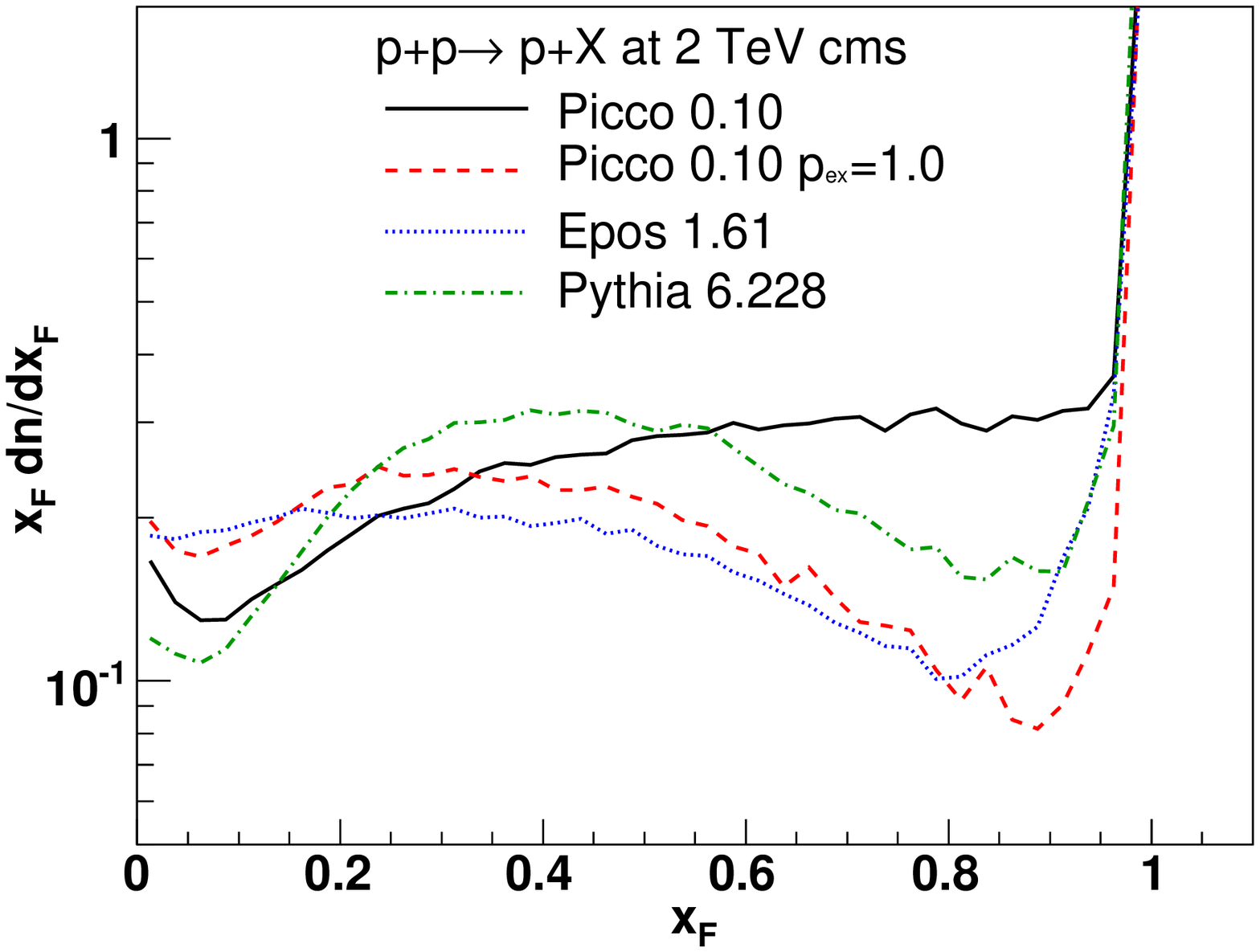}
\caption{(Color online) Feynman-x distributions of protons for
  proton-proton collisions. The data is from
  Ref. \cite{Whitmore:1973ri}. The model featuring extreme (100\%)
  remnant excitation probability, and the diquark
  approach of Pythia, lead to a dip in the forward spectrum.}
\label{fig:forward}
\end{center}
\end{figure}

In this section we consider soft remnant break-up, i.e.\ remnant
excitation. We shall first discuss the relevant parameters.  $p_{\rm
  ex}$ is the excitation probability. In the case of excitation, the
invariant mass is determined randomly according to the distribution
$dM^2/M^2$ inbetween the limits $M_{\min}$ and $M_{\max}$. In our
approach, we choose $M_{\min}$ to be the minimum mass corresponding to
the flavor content of the remnant plus twice the pion mass. In the
case of a proton-like remnant (quark content $uud$), for example,
$M_{\min}=(0.938+0.28)$~GeV. $M_{\max}$ is taken to be $0.5\sqrt{s}$. Our
simulations show that the mass limits are more of a technical nature,
their exact value does not change the results significantly. 

\begin{figure}[tb]
\begin{center}
\includegraphics[width=0.9\columnwidth]{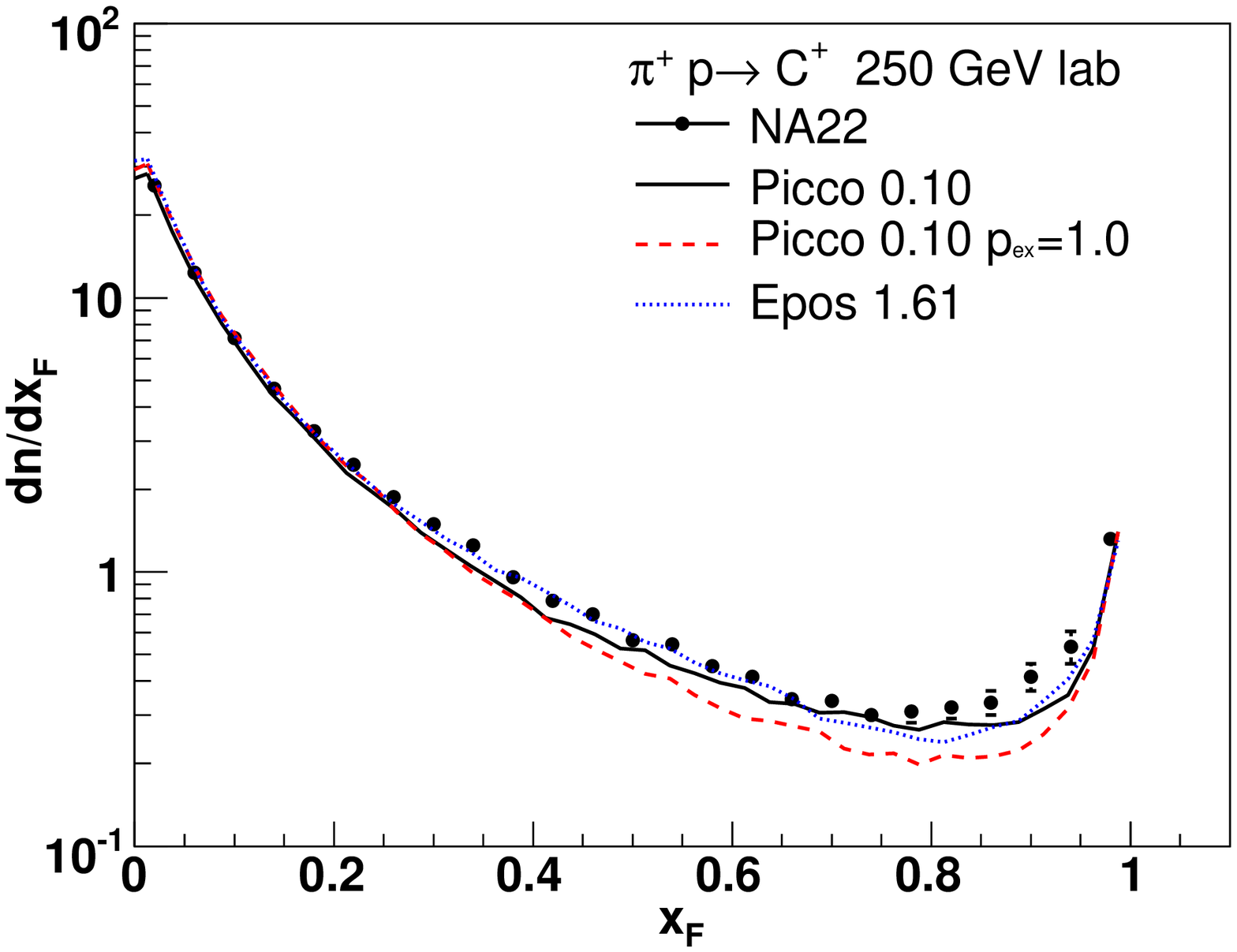}
\includegraphics[width=0.9\columnwidth]{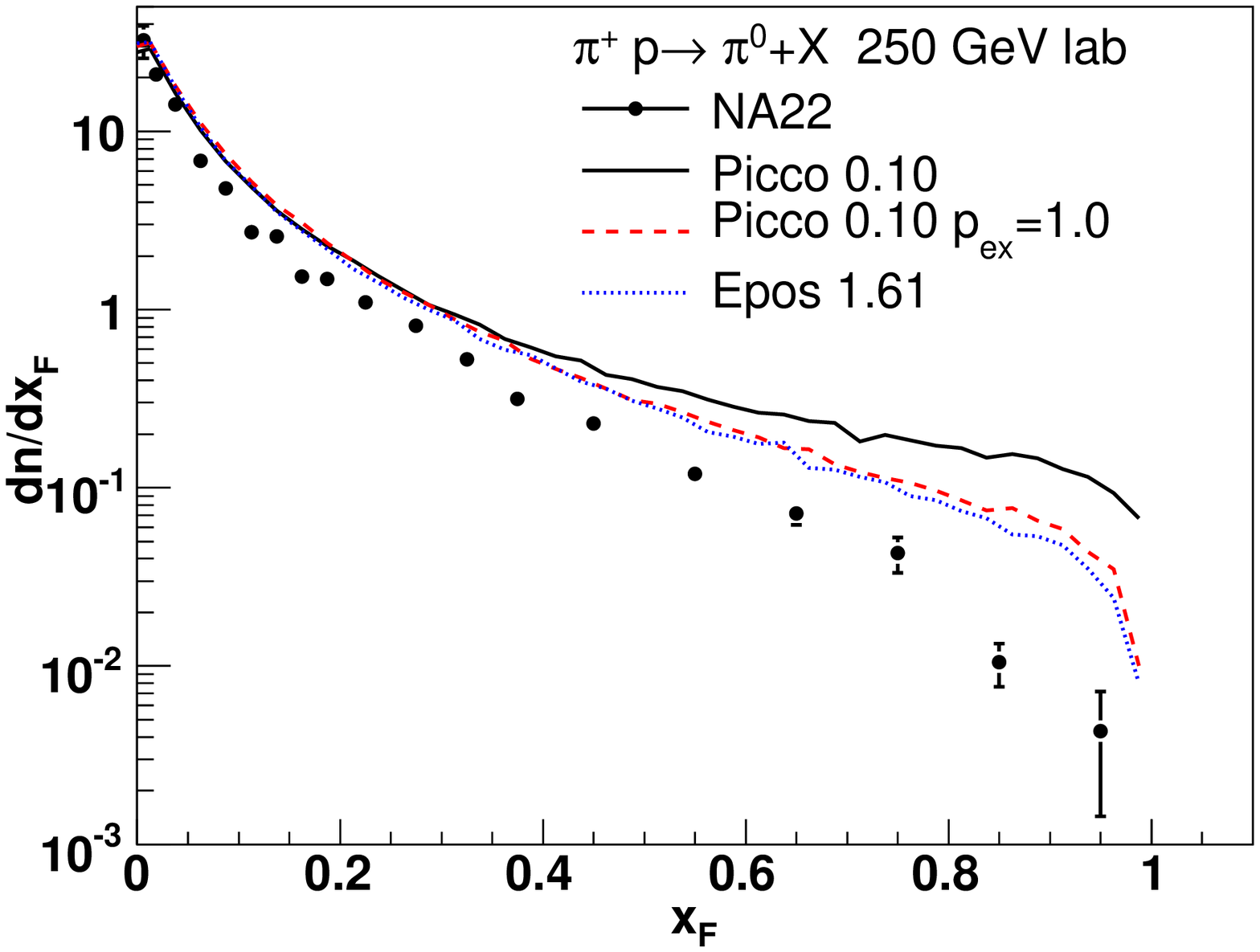}
\caption{(Color online) Feynman-x distributions of $\pi^0$ for
  $\pi^+$\-proton collisions. The data is from Ref. \cite{NA22}}
\label{fig:pi0}
\end{center}
\end{figure}

The excitation probability $p_{\rm ex}$, however, is of utmost
importance.  The exact value of this parameter has been fitted to
reproduce low-energy forward scattering data from
Ref.~\cite{Whitmore:1973ri}, see Fig.~\ref{fig:forward}. We find
$p_{\rm ex}=0.6$.  Also shown in the same figure is the forward
particle distribution for the case $p_{\rm ex}=1$, i.e., when the
remnant is excited always. Notice how this affects the forward
distribution of final-state protons: a dip appears in the region
$0.5<x_F<1$. The explanation is simple. If the remnant decays into two
or more particles, the region below $x_F<0.5$ will be populated. The
region $0.5<x_F<1$ corresponds to a mere longitudinal momentum loss of
the remnant (since it does not contain the ``participant'' quarks from
the incoming hadron) rather than to its fragmentation. The
lower panel of Fig.~\ref{fig:forward} depicts the forward particle
distribution at higher energies.

Pythia implements a different approach. Instead of treating the
remnant separately, one of the strings created in the interaction is
attached to a diquark from the projectile while the other string is
attached to the remaining valence quark. This topology, however,
appears to have problems explaining the ratio of anti-Omega to Omega
baryons. Sufficiently massive strings ending in a diquark tend to
produce more anti-baryons than baryons near that end, which is not
supported by the data \cite{Bleicher:2001nz}. In
Fig.~\ref{fig:forward} we see that a diquark string also fails to
reproduce the forward proton production data, since, similar to a decaying
remnant, the momentum of the diquark is shared among two or more
particles. 

For pion induced reactions, the situation is different, with the
remnant consisting of a simple quark-antiquark pair. Again, with a
probability $p_{\mathrm ex}$, the remnant gets excited and decays by
forming a string. A $\pi^+ + p $ reaction at 250 GeV has been studied
by the EHS/NA22 collaboration \cite{NA22}. Fig. \ref{fig:pi0} shows
the forward production of charged particles. In the case of
\textsc{Picco}, a better fit is obtained with a lower excitation
probability. \textsc{Epos}, which has a higher excitation probability
yields a even slightly better description of the data. The forward
$\pi^0$ production is shown in the lower panel of
Fig. \ref{fig:pi0}. Here, enhanced breakup improves the description of
the differential cross section, but all models are still far from the
data. In fact, no other model is able to describe this data. If one
modifies parameters extremely to reproduce neutral pions, then one
fails to describe the charged pion spectra. The situation is similar
with data on $\pi^-$ proton scattering at 360 GeV measured by the NA27
LEBC/EHS Collaboration \cite{NA27}. A new measurement of $\pi^0$
production in pion induced reactions would be extremely helpful.

\section{Influence on muon production in air showers}

How does the remnant affect muon production in air showers? The basic
reasoning is very simple: if the remnant decays into two or more
particles its energy is redistributed among particles which stay in
the hadronic channel and eventually produce muons (charged mesons,
baryons) and particles which go into the electromagnetic channel
(mainly neutral pions).

Let us first consider the break-up of baryons. The remnant of a baryon
induced collision is always a baryon. The flavor content might of
course change due to charge exchange (for example, a proton induced
reaction might give a neutron as remnant, which nevertheless remains
in the hadronic channel of the air shower). If the remnant however
gets excited and breaks up, generically it produces a number of neutral
pions. Therefore, baryonic remnant break-up should in fact reduce muon
production.

For charged-pion induced reactions, the outcome is very different.
Due to charge exchange, the remnant can be either of the flavor of a
charged pion (ignoring strangeness for the purpose of this argument)
or of a neutral pion. The decay of an excited charged pion therefore
reduces muon production since additional neutral pions are
produced. The opposite is true for the decay of neutral pion remnants,
where additional charged pions increase the energy available in the
hadronic channel.  The important question is which effect is stronger:
do charged pion remnants produce more $\pi^0$ than neutral pion
remnants produce $\pi^{\pm}$~? We start by considering low
multiplicities, when the decaying remnant produces only two
particles. With one Pomeron exchange, the remnant is with 50\%
probability charged and with 50\% probability neutral. This can be
easily seen by inserting a quark-antiquark pairs (underlined) into a
$u\bar{d}$ system:
\begin{eqnarray}
\mathrm {projectile}  &\rightarrow& \mathrm{remnant}+\mathrm{string ends}\nonumber \\
u\bar{d}  &\rightarrow&   u\underline{\bar{d}} + \underline{d}\bar{d} \label{udbar:1} \\
          &\rightarrow&   u\underline{\bar{u}} + \underline{u}\bar{d} \label{udbar:2} \\
          &\rightarrow&   \bar{d}\underline{u}+ \underline{\bar{u}}u \label{udbar:3} \\
          &\rightarrow&   \bar{d}\underline{d}+ \underline{\bar{d}}u \label{udbar:4} 
\end{eqnarray}
(\ref{udbar:1}) and (\ref{udbar:3}) will give a charged remnant, 
(\ref{udbar:2}) and (\ref{udbar:4}) a neutral one. 
The decaying remnant will try to produce an equal amount of charged
($\pi^+,\pi^-$) and neutral mesons ($\pi^0,\eta^{(\prime)} $), since
again a break-up will produce with the same probability a $u\bar{u}$
or a $d\bar{d}$-pair. But the production of $\eta^{(\prime)}$ is
suppressed due to its large mass, therefore we will find more charged
particles in the final state than neutral mesons. The consideration of
strangeness and baryon production enhances this effect.  For high
multiplicities we know that we get $1/3$ neutral and $2/3$ charged
pions, therefore the fraction of $\pi^0$ is strongly reduced.  We see
that in total, the overall number of neutral pions is reduced due to
break-up. Therefore, this mechanism will, in all, enhance muon
production.  Any additional production of strange mesons or of baryons
will amplify this effect in favor of particles which stay in the
hadronic channel of an air shower.

\begin{figure}[tb]
\begin{center}
\includegraphics[width=0.9\columnwidth]{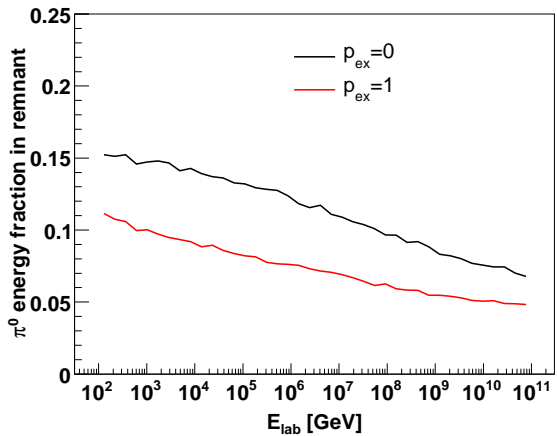}
\caption{(Color online) Energy fraction of neutral pions in the
  remnant for $\pi^+ + Air$ collisions, for the cases with and without
  break-up. Remnant break-up reduces the energy fraction in $\pi^0$
  and thus enhances muon production in air showers.}
\label{fig:xen}
\end{center}
\end{figure}

Fig.~\ref{fig:xen} shows the energy fraction of neutral pions in the
remnant, for the two extreme cases without excitation ($p_{ex}=0$) and
with complete break-up ($p_{ex}=1$).  We see that remnant break-up
indeed reduces the energy in neutral pions.

\begin{figure}[tb]
\begin{center}
\includegraphics[width=0.9\columnwidth]{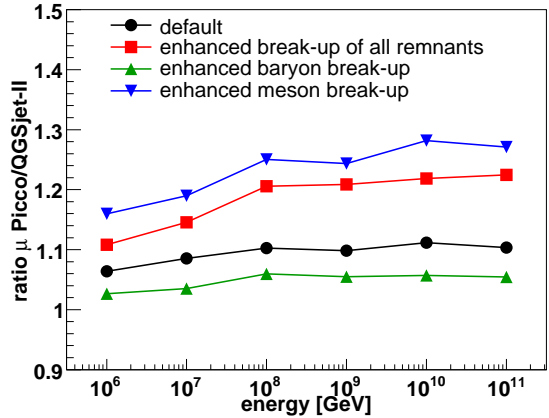}
\caption{(Color online) The ratio of total muon numbers of vertical
  proton induced air showers for enhanced break-up ($p_{\rm ex}=1.0$ instead of $p_{\rm ex}=0.6 $) of baryonic and
  mesonic remnants.}
\label{fig:muon3}
\end{center}
\end{figure}

\begin{figure}[tb]
\begin{center}
\includegraphics[width=0.9\columnwidth]{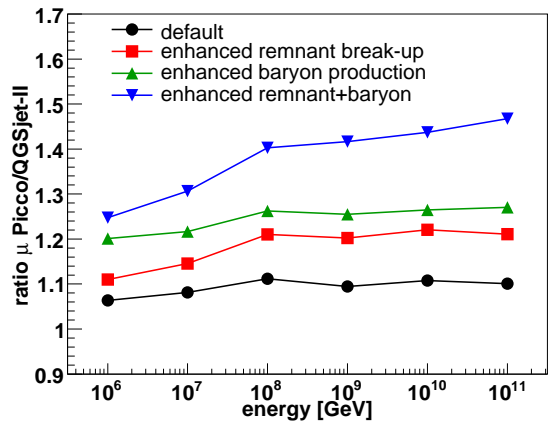}
\caption{(Color online) The ratio of total muon numbers of vertical
  proton induced air showers for different remnant break-up and baryon
  production scenarios. Enhanced baryon production means  $p_{\rm diq,str}=0.12$ and $p_{\rm diq,rem}=0.3$ instead of the default value $p_{\rm diq}=0.10$ from  \textsc{Pythia}.}
\label{fig:muon1}
\end{center}
\end{figure}

Next, we calculate the total muon number with a full air shower
simulation using \textsc{Seneca}\cite{Drescher:2002cr}. We
simulate proton induced average vertical showers, the observation
altitude is 1400~m above sea level, and we normalize to the
\textsc{QGSjet-II} model. The result is shown in Fig.~\ref{fig:muon3},
for the different cases of enhanced break-up ($p_{ex}=1$) of baryonic
and mesonic remnants.  We observe that additional break-up of baryons
reduces muon production while break-up of mesons enhances it. The net
effect is an enhancement of muons, since pion induced collisions are
more abundant.

We are now in a position to consider as well enhanced baryon
production. The parameters for the diquark production in the string
fragmentation are adjusted to give qualitatively similar results to
\textsc{EPOS}. We choose $p_{\rm diq,str}=0.12$ and $p_{\rm
  diq,rem}=0.3$ as diquark-anti-diquark pair production probabilities
in central strings and remnant strings, respectively. The higher
probability for diquark production from remnant strings is motivated
by the remnant encountering a denser target and leads to more baryons
in the forward region, as found to be important by the authors of
\textsc{EPOS}.  Fig. \ref{fig:muon1} shows that muon production in
\textsc{Picco} with default parameters (i.e.\ $p_{\rm diq,str/rem}=0.1$
from \textsc{Pythia}) is within 10\% similar to
\textsc{QGSjet-II}. Enhanced baryon production gives 25\% more
muons. However, the combined effects of both enhanced baryon
production and remnant break-up add up to 40\%.

At low energies, complete remnant break-up of baryons is excluded by
the data, (see Fig. \ref{fig:forward}) and the results shown in
Fig.~\ref{fig:muon1} should actually be considered as an upper
limit. However, we do not know whether the relevant parameter changes
at high energies. If we keep the one obtained at low energies, $p_{\rm
  ex}=0.6$, we find a rather flat distribution $x_F dn/dx_F$ of
protons at 2 TeV center of mass energy, whereas one notices a dip in
the forward scattering spectrum for the case of complete remnant
break-up.

A motivation for an enhanced remnant break-up is the fact that at
higher energies the projectile probes smaller gluon momenta in the
target and therefore encounters a higher gluon density. This effect
has already been investigated within the Color Glass Condensate (CGC)
framework for hadron-nucleus collisions at colliders in
Ref.~\cite{Dumitru:2002wd} and has been applied to air showers in
Ref.~\cite{Drescher:2004sd}. The main consequence of the enhanced
remnant break-up is a suppression of forward particle production; this
leads to a faster absorption in air showers and hence to a lower
shower maximum $X_{\rm max}$. Furthermore, an enhanced muon production
was observed in Ref. \cite{Drescher:2005ig} but was attributed mainly
to an increased overall multiplicity. Efforts are currently ongoing to
implement this particular mechanism of forward suppression into the
\textsc{Picco} model, which will allow us to test for its influence on
the muon number.  We will also investigate projectile break-up due to
a dense target in proton-proton collisions at LHC energies, and apply
this mechanism to air shower simulations.

\begin{figure}[thb]
\includegraphics[width=0.9\columnwidth]{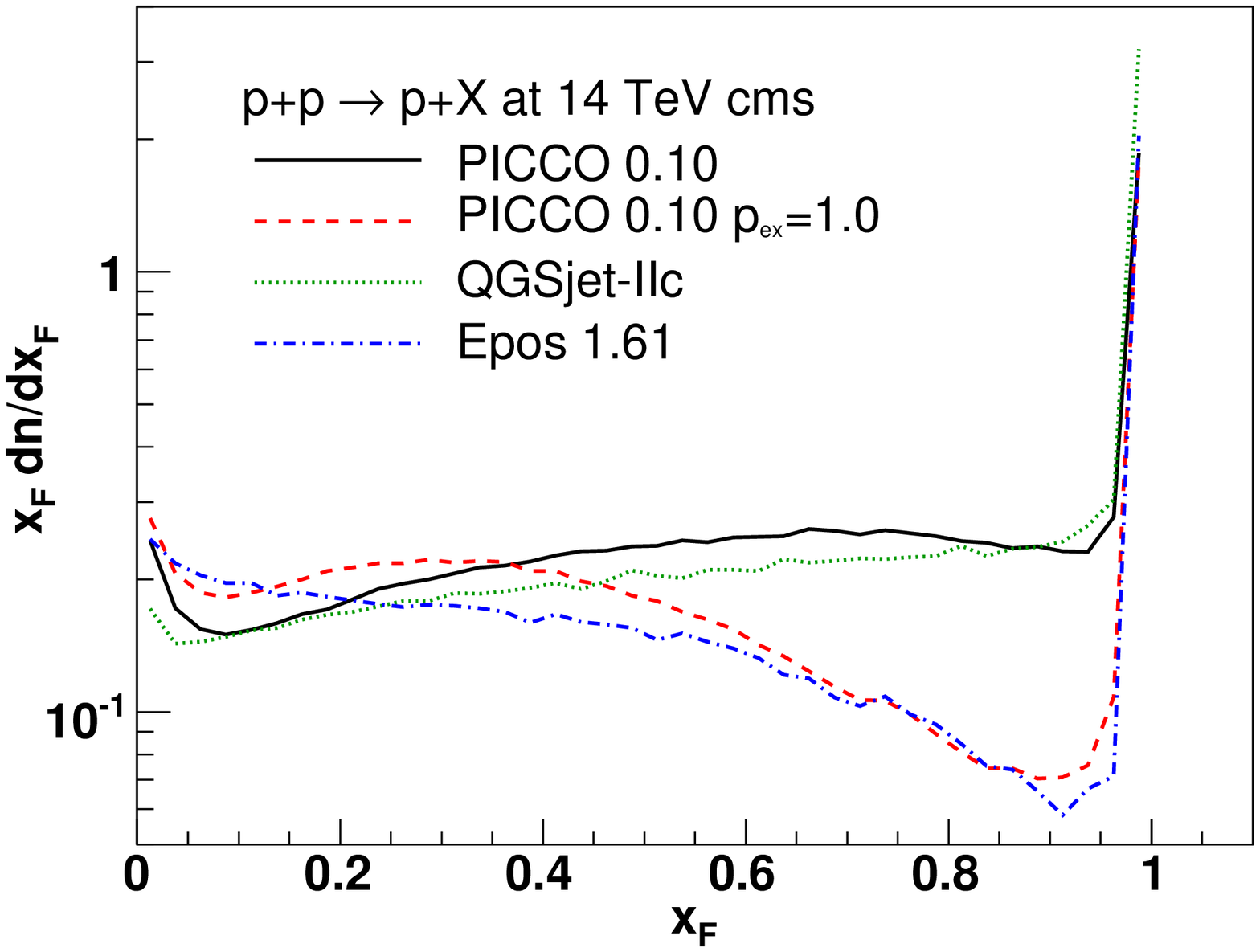}
\includegraphics[width=0.9\columnwidth]{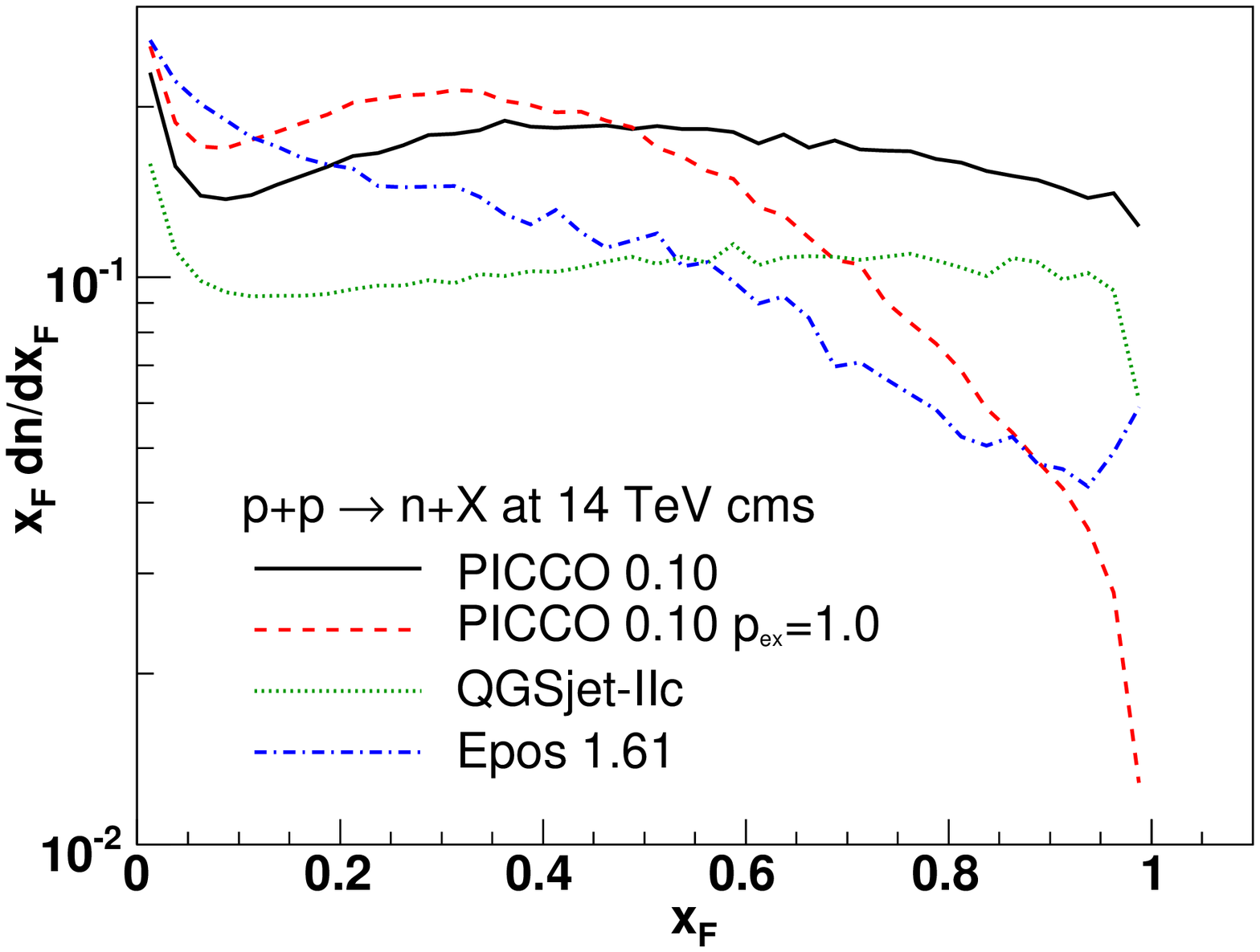}
\includegraphics[width=0.9\columnwidth]{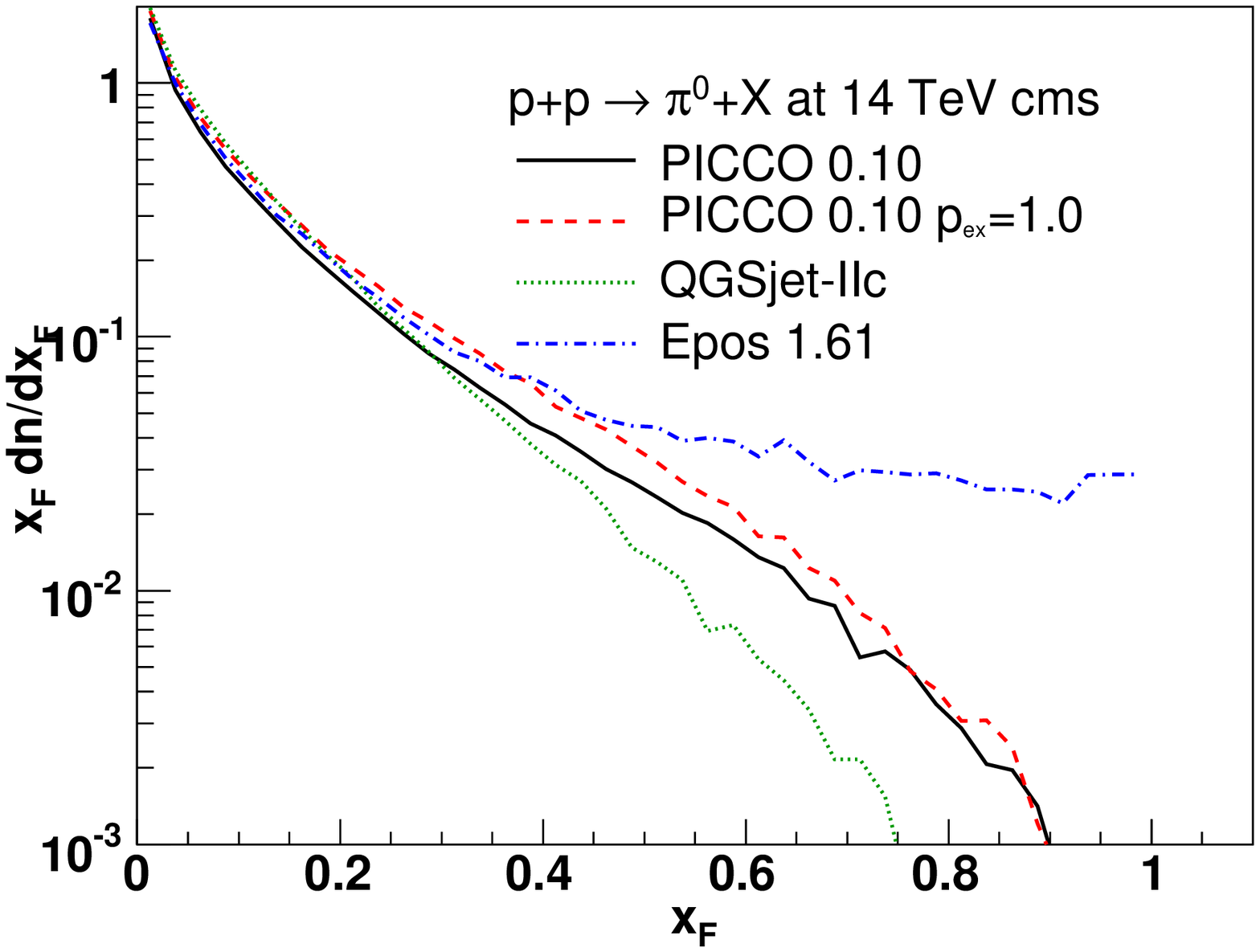}
\caption{(Color online) Feynman-$x$ distributions of identified
  particles for proton-proton collisions at 14 TeV cms energy.}
\label{fig:mult}
\end{figure}

\section{Forward scattering at LHC energies}

Finally, we compare model predictions in the kinematic range relevant
for the forthcoming LHCf \cite{LHCf} experiment.  This experiment
plans to measure photons, neutral pions and neutrons in the very
forward region with a small detector between the two beam pipes of the
LHC collider. Except for the diffractive peak, neutrons and protons
are expected to show similar distributions in this region, with
neutrons being somewhat outnumbered. The results are shown in
Fig.~\ref{fig:mult}.  The slope of the forward neutron spectrum will
give insight into remnant break-up at these energies and will
therefore allow one to constrain muon production in air showers. It is
also interesting to examine the absolute normalization of the very
forward neutron spectrum as this relates directly to the treatment of
the flavor content of the remnant. \textsc{QGSjet-II} predicts less
neutrons since it allows for only one charge exchange reaction. 

As for neutral pions: here the differences due to remnant break-up are much
smaller. The relatively flat forward spectrum of \textsc{EPOS} stems
from a specific implementation of the popcorn effect \cite{popcorn},
where the two leading particles of string fragmentation can be
interchanged with some probability.

\section{Summary}

We have shown that remnant fragmentation in hadronic interaction
models for air showers has significant influence on muon
production. If the projectile remnant decays into two or more
particles the energy fraction which goes into the hadronic or
electromagnetic channel changes. Enhanced baryon break-up reduces and
enhanced meson break-up increases muon production in air showers.  At
low energies, forward scattering data constrains the break-up
probability of the remnant to roughly 60\%.  The measurements of the
LHCf experiment will be extremely helpful for our understanding of
remnant break-up at higher energies, and hence will provide important
constraints for muon production in air showers.

\begin{acknowledgments}
 This work is supported by BMBF grant 05~CU5RI1/3. HJD thanks
 S.Ostapchenko, A.Dumitru, T. Pierog, K.Werner, R.Engel and G.Farrar
 for useful comments.
\end{acknowledgments}


%
\end{document}